\documentclass[rmp,twocolumn,showpacs,superscriptaddress]{revtex4}

\usepackage{dcolumn,graphicx,amsmath,amssymb,txfonts}

\newcommand\av[1]{\bar{#1}}
\newcommand\defn{\textit}
\newcommand{\Ord}{\mathop{\mathrm{O}}\nolimits}

\begin{document}

\title{Nonequilibrium phase transition in the coevolution of networks and
opinions}

\author{Petter Holme}
\affiliation{Department of Computer Science, University of New Mexico,
  Albuquerque, NM 87131, U.S.A.}
\affiliation{Department of Physics, University of Michigan,
  Ann Arbor, MI 48109, U.S.A.}
\author{M. E. J. Newman}
\affiliation{Department of Physics, University of Michigan,
  Ann Arbor, MI 48109, U.S.A.}

\begin{abstract}
  Models of the convergence of opinion in social systems have been the
  subject of a considerable amount of recent attention in the physics
  literature.  These models divide into two classes, those in which
  individuals form their beliefs based on the opinions of their neighbors
  in a social network of personal acquaintances, and those in which,
  conversely, network connections form between individuals of similar
  beliefs.  While both of these processes can give rise to realistic levels
  of agreement between acquaintances, practical experience suggests that
  opinion formation in the real world is not a result of one process or the
  other, but a combination of the two.  Here we present a simple model of
  this combination, with a single parameter controlling the balance of the
  two processes.  We find that the model undergoes a continuous phase
  transition as this parameter is varied, from a regime in which opinions
  are arbitrarily diverse to one in which most individuals hold the same
  opinion.  We characterize the static and dynamic properties of this
  transition.
\end{abstract}


\maketitle

\section{Introduction}

Simple mathematical models describing emergent phenomena in human
populations~\cite{schelling}, such as voter models and market models, have
a long history of study in the social sciences.  It is only relatively
recently, however, that physicists have noted the close conceptual and
mathematical connections between these models and traditional models in
statistical physics such as spin models.  Building on this observation,
there have been a number of important advances in the understanding of
these models in the last decade or so, most notably in the study of social
networks~\cite{AB02,DM02,Newman03d}.  While the physics community has been
concerned primarily with studies of network structure, there has also been
a substantial line of investigation focusing on dynamical processes on
networks.  One example, which has a long history in sociology but is also
well suited to study using physics methods, is the dynamics of opinion
formation.  This problem highlights one of the fundamental questions in
network dynamics, namely whether dynamics controls the structure of a
network or the structure controls the dynamics.

It is observed that real social networks tend to divide into groups or
communities of like-minded individuals.  An obvious question to ask is
whether individuals become like-minded because they are connected via the
network~\cite{liggett:vot,cas:vot,sood:vot,sznajd,deuff:opi,cc:axe}, or
whether they form network connections because they are
like-minded~\cite{mcp:bird}.  Both situations have been studied with
physics-style models, the first using opinion formation
models~\cite{liggett:vot,cas:vot,sood:vot} and the second using models of
``assortative mixing'' or
``homophily''~\cite{Soderberg02,Newman03c,BPDA04}.  Common sense, however,
tells us that the distinction between the two scenarios is not clear-cut.
Rather, the real world self-organizes by a combination of the two, the
network changing in response to opinion and opinion changing in response to
the network.  In this paper we study a simple model---perhaps \emph{the}
simplest---that combines opinion dynamics with assortative network
formation, revealing an apparent phase transition between regimes in which
one process or the other dominates the dynamics.

\section{Model definition}

Consider a network of $N$ vertices, representing individuals, joined in
pairs by $M$ edges, representing acquaintance between
individuals\footnote{
  Although acquaintance networks are typically simple graphs, with
  multiedges and self-edges disallowed, we have in the interest of
  simplicity, allowed multiedges and self-edges in our calculation.  Since
  these form only a small fraction of all edges, we expect that our results
  would change little if we were to remove them.}.  Each individual is assumed to hold one of~$G$
possible opinions on some topic of interest.  The opinion of individual~$i$
is denoted~$g_i$.  In the past, researchers have considered both cases
where $G$ is a fixed small number, such as a choice between candidates in
an election~\cite{cas:vot,sood:vot,sznajd}, and cases in which the number
of possible opinions is essentially unlimited~\cite{deuff:opi}, so that $G$
can be arbitrarily large.  An example of the latter might be religious
belief (or lack of it)---the number of subtly different religious beliefs
appears to be limited only by the number of people available to hold them.

The case of fixed small~$G$ has relatively simple behavior compared to the
case of arbitrarily large~$G$, and so it is on the latter that we focus
here.  We will assume that the number of possible opinions scales in
proportion to the number of individuals, and parameterize this
proportionality by the ratio $\gamma=N/G$.  (It is possible that not all
opinions will end up existing in the population.  Our model allows for some
opinions to become extinct as the dynamics evolves, so that the final
number of distinct opinions may be less than~$G$.)

\begin{figure}
  \resizebox*{.9\linewidth}{!}{\includegraphics{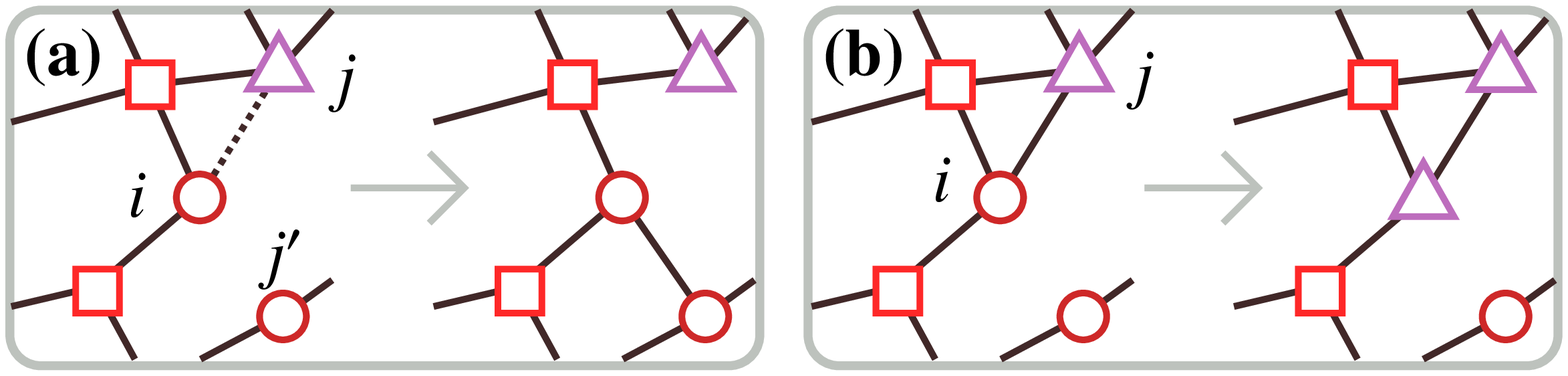}}
  \caption{An illustration of our model, with vertex shapes representing
  opinions.  At each time step the system is updated according to the
  process illustrated in panel~(a) with probability~$\phi$ or panel~(b)
  with probability $1-\phi$.  In (a) a vertex $i$ is selected at random and
  one of its edges (in this case the edge $(i,j)$) is rewired to a new
  vertex $j'$ holding the same opinion as~$i$.  In (b) vertex~$i$ adopts
  the opinion of one of its neighbors~$j$.
  }
  \label{fig:ill}
\end{figure}

The $M$ edges of the network are initially placed uniformly at random
between vertex pairs, and opinions are assigned to vertices uniformly at
random.  We then study by computer simulation a dynamics in which on each
step of the simulation we either move an edge to lie between two
individuals whose opinions agree, or we change the opinion of an individual
to agree with one of their neighbors.  To be specific, on each step we do
the following (see Fig.~\ref{fig:ill}).
\begin{enumerate}
\item\label{step:rewire} Pick a vertex~$i$ at random.  If the degree
$k_i$ of that vertex is zero, do nothing.  Otherwise, with
probability~$\phi$, select at random one of the edges attached to $i$ and
move the other end of that edge to a vertex chosen randomly from the set of
all vertices having opinion~$g_i$.
\item\label{step:vote} Otherwise (i.e.,~with probability $1-\phi$) pick a
random neighbor~$j$ of $i$ and set $g_i$ equal to~$g_j$.
\end{enumerate}
Step~\ref{step:rewire} represents the formation of new acquaintances
between people of similar opinions. Step~\ref{step:vote} represents the
influence of acquaintances on one another, opinions becoming similar as a
result of acquaintance.

Note that both the total number of edges~$M$ in our network and the total
number of possible opinions~$G$ are fixed.  In the limit of large system
size, the model thus has three parameters: the average degree
$\av{k}=2M/N$, the mean number of people holding an opinion~$\gamma=N/G$,
and the parameter~$\phi$.  In our studies, we primarily keep the first two
of these parameters fixed and ask what happens as we vary the third.

\section{Phases and critical scaling of community sizes}

The expected qualitative behavior of the model is clear.  Since both of our
update moves tend to decrease the number of nearest-neighbor vertex pairs
with different opinions, we should ultimately reach a state in which the
network is divided into a set of separate components, disconnected from one
another, with all members of a component holding the same opinion.  That
is, the model segregates into a set of communities such that no individual
has any acquaintances with whom they disagree.  We call this the
\defn{consensus state}.  Furthermore, once we reach consensus, all moves in
the model involve the random rearrangement of edges within components, and
hence, in the limit of long time, the components become random graphs
with uniform uncorrelated arrangements of their edges.

The primary interest in our model therefore is in the number and sizes of
the communities that form and in the dynamics of the model as it comes to
consensus.  Let us consider the distribution $P(s)$ of the sizes~$s$ of the
consensus communities.  In the limit $\phi\to1$, only updates that move
edges are allowed and hence the consensus state is one in which the
communities consist of the sets of initial holders of the individual
opinions.  Since the initial assignment of opinions is random, the sizes of
these sets follow the multinomial distribution, or the Poisson distribution
with mean~$\gamma$ in the limit of large~$N$.  Conversely, in the limit
$\phi\to0$, only changes of opinion are allowed and not edge moves, which
means that the communities correspond to the initial components in the
graph, which are simply the components of a random graph.  Assuming we are
in the regime $\av{k}>1$ in which a giant component exists in the random
graph, we will then have one giant (extensive) community and an exponential
distribution of small communities.  Thus, in varying~$\phi$ we go from a
situation in which we have only small communities with constant average
size~$\gamma$ to one in which we have a giant community plus a set of small
ones.

This is the classic behavior seen in a system undergoing a continuous phase
transition and it leads us to conjecture that our model displays a phase
transition with decreasing $\phi$ at which a giant community of like-minded
individuals forms.  In other words, there is a transition between a regime
in which the population holds a broad variety of views and one in which
most people believe the same thing.  We now offer a variety of further
evidence to support this conjecture.  (Phase transition behavior is also
seen in some models of opinion formation on static networks, such as the
model of Ref.~\cite{cc:axe}, although the mechanisms at work appear to be
different from those considered here.)

\begin{figure}
  \resizebox*{0.9\linewidth}{!}{\includegraphics{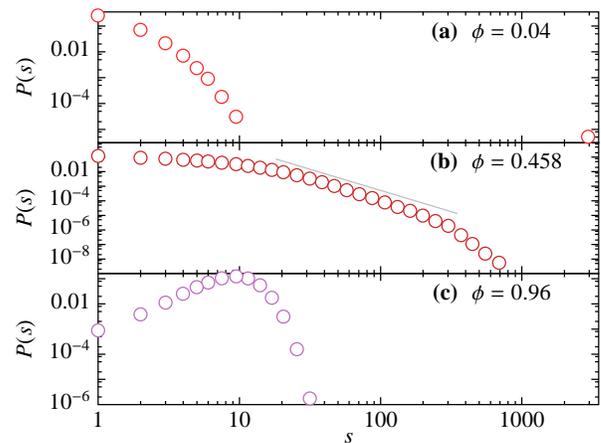}}
  \caption{Histograms of community sizes in the consensus state for values
  of $\phi$ above, at, and below the critical point in panels (a), (b), and
  (c) respectively.  Values of the other parameters are $N=3200$, $M=6400$
  (giving $\av{k}=4$), and $\gamma=10$.  In panel~(b) the distribution
  appears to follow a power law for part of its range with exponent $3.5\pm
  0.3$, as indicated by the solid line.  Numerical data are averaged over
  $10^4$ realizations for each $\phi$-value and binned logarithmically.
  }
  \label{fig:hist}
\end{figure}

In Fig.~\ref{fig:hist} we show plots of $P(s)$ from simulations of our
model for $\av{k}=4$ and $\gamma=10$.  As the figure shows, we do indeed
see a qualitative change from a regime with no giant community to one with
a giant community.  At an intermediate value of $\phi$ around $0.458$ we
find a distribution of community sizes that appears to follow a power law
$P(s)\sim s^{-\alpha}$ over a significant part of its range, another
typical signature of criticality.  The exponent~$\alpha$ of the power law
is measured to be $3.5\pm0.3$, which is incompatible with the value $2.5$
of the corresponding exponent for the phase transition at which a giant
component forms in a random graph (a transition which belongs to the
mean-field percolation universality class).

\begin{figure}
  \resizebox*{0.9\linewidth}{!}{\includegraphics{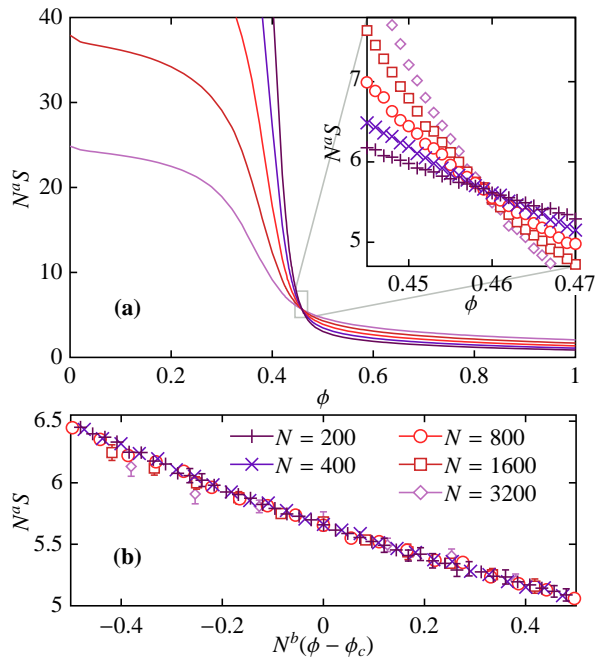}}
  \caption{Finite size scaling analysis for $\av{k}=4$ and $\gamma=10$.
  (a)~Crossing plot used to determine the critical point~$\phi_c$ and
  exponent~$a$.  We find $\phi_c=0.458\pm 0.008$ and $a=0.61\pm 0.05$.  The
  inset shows a blow-up of the region around the critical point.
  (b)~Scaling collapse used to determine the exponent~$b$ which is found to
  take the value $b=0.7\pm 0.1$.  The data are averaged over $10^4$
  realizations for $\phi$-value.  Error bars are shown where they are larger
  than the symbol size.
  }
  \label{fig:fss}
\end{figure}

To further investigate our transition, we perform a finite size scaling
analysis in the critical region.  To do this, we need first to choose an
order parameter for the model.  The obvious choice is the size~$S$ of the
largest community in the consensus state as a fraction of system size.  The
arguments above suggest that this quantity should be of size $\Ord(N^{-1})$
for values of $\phi$ above the phase transition (and hence zero in the
thermodynamic limit) and $\Ord(1)$ below it.  We assume a scaling relation
of the form
\begin{equation}\label{eq:fss}
  S = N^{-a}\,F\Bigl(N^b(\phi-\phi_c)\Bigr),
\end{equation}
where $\phi_c$ is the critical value of $\phi$ (which is presumably a
function of $\av{k}$ and~$\gamma$), $F$~is a universal scaling function
(bounded as its argument tends to $\pm\infty$), and $a$ and $b$ are
critical exponents.  To estimate $\phi_c$ we plot $ N^aS$ against $\phi$
and tune $a$ such that the results for simulations at different~$N$ but
fixed $\av{k}$ and $\gamma$ cross at a single point, which is the critical
point.  Such a plot for $\av{k}=4$ and $\gamma=10$ is shown in
Fig.~\ref{fig:fss}(a).  With $a=0.61\pm0.05$ we obtain a unique crossing
point at $\phi_c=0.458\pm0.008$, which agrees well with the previous rough
estimate of $\phi_c$ from Fig.~\ref{fig:hist}.

Using this value we can now determine the exponent $b$ by plotting $N^aS$
against $N^b(\phi-\phi_c)$.  Since $F(x)$ is a universal function, we
should, for the correct choice of~$b$, find a data collapse in the critical
region.  In Fig.~\ref{fig:fss}(b) we show that such a data collapse does
indeed occur for $b=0.7\pm0.1$.

\begin{figure}
  \resizebox*{0.9\linewidth}{!}{\includegraphics{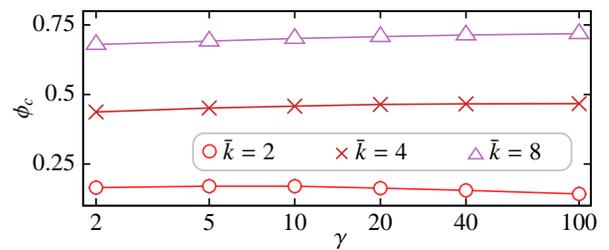}}
  \caption{Values of $\phi_c$ as a function of $\gamma$ for various
  $\av{k}$ obtained by finite size scaling analyses using system sizes
  $N=200$, $400$, $800$, and $1600$ and $10^4$ realizations for each size
  and set of parameter values.  Note that the horizontal axis is
  logarithmic.
  }
  \label{fig:phd}
\end{figure}

We have performed similar finite size scaling analyses for a variety of
other points $(\av{k},\gamma)$ in the parameter space and, as we would
expect, we find that the position $\phi_c$ of the phase transition
varies---see Fig.~\ref{fig:phd}---but that good scaling collapses exist at
all parameter values for values of the critical exponents consistent with
the values $a=0.61$ and $b=0.7$ found above.

Despite the qualitative similarities between the present phase transition
and the percolation transition, our exponent values for $a$ and $b$ show
that the two transitions are in different universality classes: the
corresponding exponents for random graph percolation are $a=b=\frac13$,
which are incompatible with the values measured above.

\begin{figure}[b]
  \resizebox*{\linewidth}{!}{\includegraphics{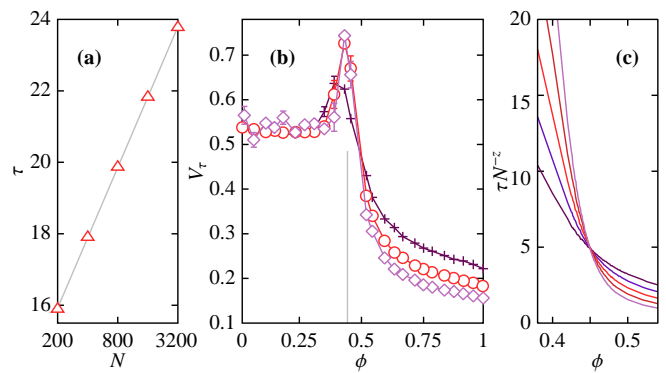}}
  \caption{Scaling of the average time $\tau$ to reach consensus.
  (a)~Convergence time as a function of system size for $\phi=1$.  The
  straight line is a fit to a logarithmic form and indicates
  that~$\tau\sim\log N$.  (b)~Coefficient of variation of the convergence
  time as a function of~$\phi$.  The vertical gray line marks the position
  of the critical point $\phi_c=0.458$.  (c)~Scaling plot used to determine
  the dynamical exponent $z$.  The crossing point falls at $\phi = 0.45\pm
  0.02$ consistent with the value of $\phi_c$ obtained above.  The
  dynamical exponent is found to take the value $z=0.61\pm 0.15$.  Parameter
  values are $\av{k}=4$ and $\gamma=10$ in all panels.  All data points are
  averaged over $10^4$ realizations.  Symbols are the same as in
  Fig.~\ref{fig:fss}.  For the sake of clarity, system sizes $N=400$ and
  $N=1600$ are omitted in~(b).
  }
  \label{fig:time}
\end{figure}

\section{Dynamical critical behavior}

Our model differs from percolation in another important respect also:
percolation is a static, geometric phase transition, whereas the present
model is fundamentally dynamic, the consensus arising as the limiting fixed
point of a converging non-equilibrium dynamics.  It is interesting
therefore to explore the way in which our model approaches consensus.

In previous studies of opinion formation models of this type on fixed
networks a key quantity of interest is the average convergence time~$\tau$,
which is the number of updates per vertex needed to reach consensus.  If
$\phi=0$ then $\tau$ is known to scale as $\tau\sim N$ as system size
becomes large~\cite{sood:vot}.  In the opposite limit ($\phi=1$), opinions
are fixed and convergence to consensus involves moving edges one by one to
fall between like-minded pairs of individuals.  This is a standard
sampling-with-replacement process in which the number $U$ of unsatisfied
edges is expected to decay as $U \sim Me^{-t/M}$ for large times~$t$.  Thus
the time to reach a configuration in which $U=\Ord(1)$ is $t\sim M\log M$,
and the convergence time is this quantity divided by the system size~$N$.
For fixed average degree $\av{k}=2M/N$, this then implies that
$\tau\sim\log N$.  This result is confirmed numerically in
Fig.~\ref{fig:time}(a).

For $\phi$ close to~$\phi_c$, experience with other phase transitions leads
us to expect critical fluctuations and critical slowing down in~$\tau$.
Figure~\ref{fig:time}(b) shows that indeed there are large fluctuations in
the convergence time in the critical region.  The figure shows the value of
the coefficient of variation~$V_\tau$ of the consensus time (i.e.,~the
ratio of the standard deviation of $\tau$ to its mean) as a function of
$\phi$ and a clear peak is visible around $\phi_c\simeq0.46$.  To
characterize the critical slowing down we assume that $\tau$ takes the
traditional scaling form $\tau\sim N^z$ at the critical point, where $z$ is
a dynamical exponent~\cite{katya:swxy}.  Figure~\ref{fig:time}(c) shows a
plot of $\tau N^{-z}$ as a function of~$\phi$.  If the system follows the
expected scaling at $\phi_c$ then the resulting curves should cross at the
critical point.  Although good numerical results are considerably harder to
obtain in this case than for the community sizes presented earlier, we find
that the curves cross at a single point if $z=0.61\pm0.15$ and
$\phi=0.44\pm0.03$, the latter being consistent with our previous value of
$\phi_c=0.46$ for the position of the phase transition.

\section{Summary and conclusions}

To summarize, we have proposed a simple model for the simultaneous
formation of opinions and social networks in a situation in which both
adapt to the other.  Our model contrasts with earlier models of opinion
formation in which social structure is regarded as static and opinions are
an outcome of that pre-existing
structure~\cite{cc:axe,bik:fads,arthur:contagion,watts:fad,our:fad2}.
Our model is a dynamic, non-equilibrium model that reaches a consensus state in
finite time on a finite network.  The structure of the consensus state
displays clear signatures of a continuous phase transition as the balance
between the two processes of opinion change and network rewiring is varied.
We have demonstrated a finite size scaling data collapse in the critical
region around this phase transition, characterized by universal critical
exponents independent of model parameters.  The approach to the consensus
state displays critical fluctuations in the time to reach consensus and
critical slowing down associated with an additional dynamical exponent.
The phase transition in the model is of particular interest in that it
provides an example of a simple process in which a fundamental change in
the social structure of the community can be produced by only a small
change in the parameters of the system.

Finally, we note that for the specific example of opinion formation
mentioned in the introduction---that of choice of religion---it is known
that the sizes of the communities of adherents of religious sects are in
fact distributed, roughly speaking, according to a power law~\cite{ZM01}.
This may be a signature of critical behavior in opinion formation, as
displayed by the model described here, although other explanations, such as
the Yule process~\cite{Yule25,Simon55}, are also possible.

\begin{acknowledgements}
The authors thank Paul Krapivsky and Claudio Castellano for helpful
suggestions and comments. This work was supported in part by the
Wenner-Gren Foundations (PH) and the National Science Foundation under
grant number DMS--0234188 (MEJN).
\end{acknowledgements}

\end{document}